\documentclass[runningheads]{llncs}
\usepackage{amsmath}
\usepackage{multirow}
 \usepackage{tabularx}
\usepackage{array}
\newcolumntype{x}[1]{>{\centering\arraybackslash\hspace{0.1pt}}m{#1}}
\usepackage{graphicx}
\usepackage{pgfplots}
\usepackage[all]{nowidow}
\usepackage[utf8]{inputenc}
\usepackage{tikz}
\usetikzlibrary{er,positioning,bayesnet}
\usepackage{multicol}
\usepackage{algpseudocode,algorithm,algorithmicx}
\usepackage[a4paper, total={5.5in, 9in},centering]{geometry}
\begin{document}

\title{Complete characterization of multi-channel single~photon ionization}
%
%
\author{Jasper Peschel\inst{1}$^\text{,*}$ \and
David Busto\inst{1,2}$^\text{,*}$ \and
Marius Plach\inst{1} \and
Mattias Bertolino\inst{1} \and
Maria Hoflund\inst{1} \and
Sylvain Maclot\inst{1}$^\text{,\dag}$ \and
Jimmy Vinbladh\inst{1,3} \and
Hampus Wikmark\inst{1}$^\text{,\ddag}$  \and
Felipe Zapata\inst{1} \and
Eva Lindroth\inst{3} \and
Mathieu Gisselbrecht\inst{1} \and
Jan Marcus Dahlström\inst{1} \and
Anne L'Huillier\inst{1} \and
Per Eng-Johnsson\inst{1}}

\authorrunning{J. Peschel et al.}

\institute{Department of Physics, Lund University, P.O. Box 118, 22100 Lund, Sweden\\ \and
Physikalisches Institut, Albert-Ludwigs-Universität, Stefan-Meier-Strasse 19, 79104 Freiburg, Germany\\ \and
Department of Physics, Stockholm University, AlbaNova University Center, SE-106 91 Stockholm, Sweden\\
$^\text{\dag}$~Current address: Department of Physics, Gothenburg University, Box 100, 405 30, Gothenburg, Sweden\\
$^\text{\ddag}$~Current address: Department of Physics and Astronomy, Uppsala University, Box 516, 751 20, Uppsala, Sweden\\
$^\text{*}$~These authors contributed equally to this work\\
}
\maketitle              
\begin{abstract}
Ionization of atoms and molecules by absorption of a light pulse results in electron wavepackets carrying information on the atomic or molecular structure as well as on the dynamics of the ionization process. These wavepackets can be described as a coherent sum of waves of given angular momentum, called partial waves, each characterized by an amplitude and a phase. The complete characterization of the individual angular momentum components is experimentally challenging, requiring the analysis of the interference between partial waves both in energy and angle. Using a two-photon interferometry technique based on extreme ultraviolet attosecond and infrared femtosecond pulses, we characterize the individual partial wave components in the photoionization of the $2p^6$ shell in neon. The study of the phases of the angular momentum channels allows us to unravel the influence of short-range, correlation and centrifugal effects. This approach enables the complete reconstruction of photoionization electron wavepackets in time and space, providing insight into the photoionization dynamics. 

\end{abstract}
%
%
%
Photoionization is a fundamental process that happens when electromagnetic radiation of high enough frequency is absorbed by matter. Since the 70s, synchrotron radiation has been used for photoionization studies
\cite{KennedyPRA1972,SchmidtRPP1992,BeckerJESRP1998}, playing an important role for our understanding of the quantum nature of matter and its interaction with light. In general, an electron is released into the continuum via various ionization channels corresponding to different initial or final angular momenta. In these channels, the probability to find the released electron in space is proportional to the square of a spherical harmonic function, as exemplified in Figure 1 for the case of the ejection of an electron from the $2p^6$ shell of neon. The characterization of the photoionization process  requires the determination of the amplitudes of the different channels and, when they add coherently, their relative phases.
Experimentally this requires angular detection, to analyse the interference between different outgoing angular momentum channels \cite{CooperJCP1968,CooperPRA1993}. Additionally,  
energy resolution is needed to disentangle ionization pathways leading to different ionic states or distinguish between different excitation schemes.

In order to perform ``complete'' experiments, where the outgoing electron wavepacket is reconstructed, additional experimental information is usually needed \cite{BeckerJESRP1998,BedersonCAMP1969,EminyanPRL1973}. 
Advances in this direction have been made by preparing atoms in aligned or oriented excited states, using light with several frequency and polarization components~\cite{GodehusenPRA1998,CherepkovJESRP2005,WuilleumierJPB2006}. Recent experiments have extended such studies to ground state atoms using two-photon schemes combining either laser-generated attosecond extreme ultraviolet (XUV) pulses~\cite{PaulScience2001,HaberPCA2009} or free-electron laser femtosecond pulses~\cite{MazzaNatComm2014} with infrared (IR) laser pulses. These have been limited, however, to the characterization of the final two-photon wavepacket~\cite{KlunderPRL2011,HaberPCA2009,MazzaNatComm2014}.

\begin{figure}[t!]
  \centering
  \includegraphics[width=0.8\textwidth]{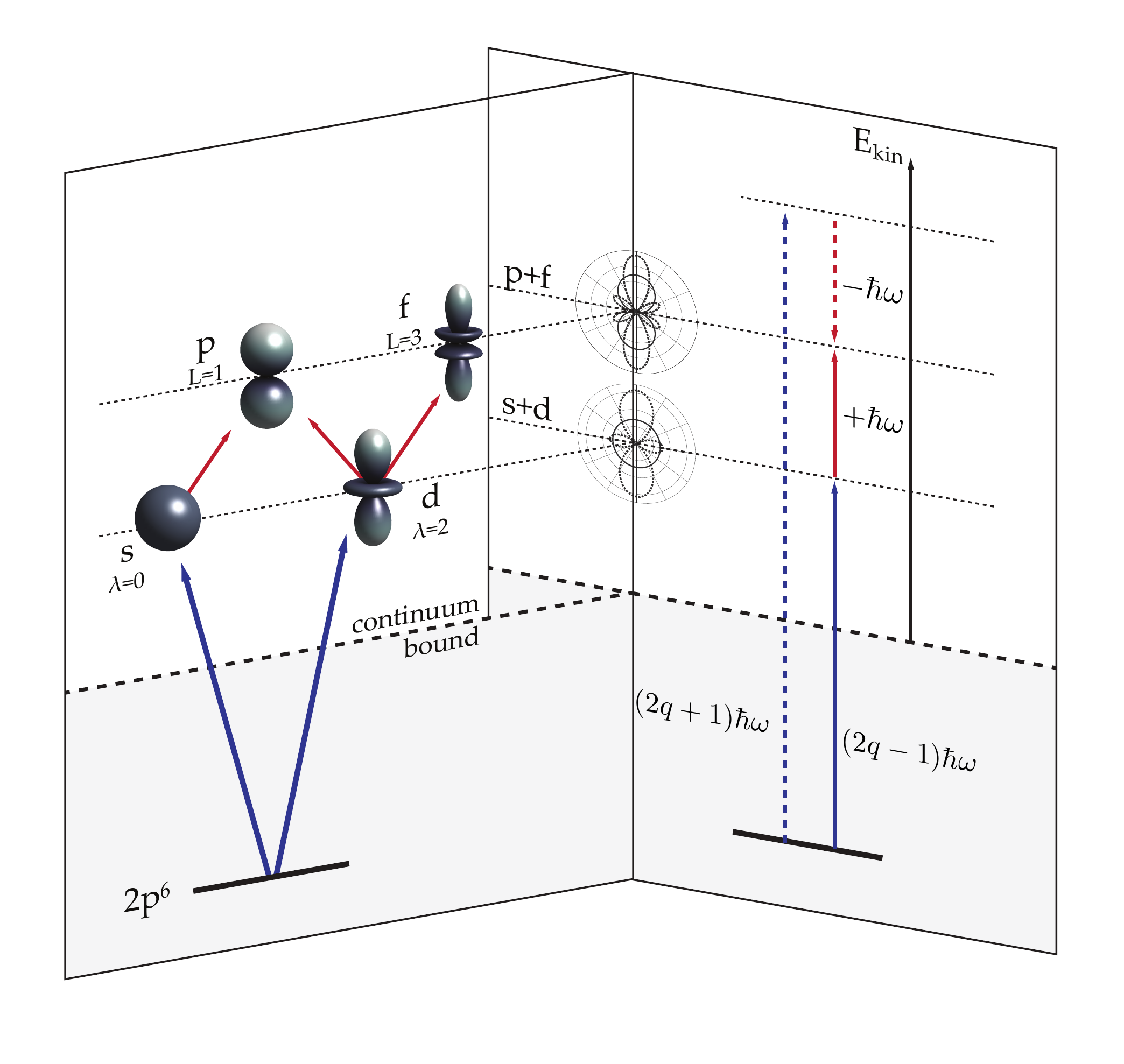}
  \caption{\label{pathways}\textbf{Angular momenta and energy level diagram:} Angular channels involved in the one- and two-photon transitions addressed in the present work (left panel) and principle of the interferometric RABBIT technique (right panel). The photoionization from the $2p^6$-ground state of neon towards $s$- and $d$- continuum states (blue arrows) is followed by the absorption or stimulated emission of an additional IR photon (red arrows, solid and dashed lines, right panel), leading to final $p$- and $f$-states. The left panel presents the quantum paths corresponding to the initial state $m=0$. For $m=\pm 1$, the path going via $\lambda=0$ is forbidden.}
\end{figure}

In this article, we completely characterize one-photon ionization from the $2p^6$-ground state of neon, obtaining the amplitude and relative phase of the electric dipole transition matrix elements towards $s$- and $d$-continuum states. To extract channel-resolved one-photon ionization (scattering) phases, we apply the reconstruction of attosecond beating by interference of two-photon transitions (RABBIT) technique \cite{PaulScience2001} while detecting electrons with full angular resolution \cite{AseyevPRL2003,Heuser2016,VilleneuveScience2017,Cirelli2018}. We use XUV pulses from high-order harmonic generation (HHG)~\cite{FerrayJPB1988} in argon in the 20-50 eV range and delayed weak IR pulses from the laser driving the HHG process. The phase extraction relies on the fact that transitions between continuum states are universal, \textit{i.e.} independent of the atom. We then determine the channel-resolved one-photon amplitudes from measurements with only the XUV field.  The retrieved values for the scattering phases and channel amplitudes are in excellent agreement with calculations using angular-channel-resolved many-body perturbation theory \cite{DahlstromPRA2012,DahlstromJPB2014,VinbladhPRA2019}.


\begin{figure}[t!]
  \centering
     \includegraphics[width=0.8\textwidth]{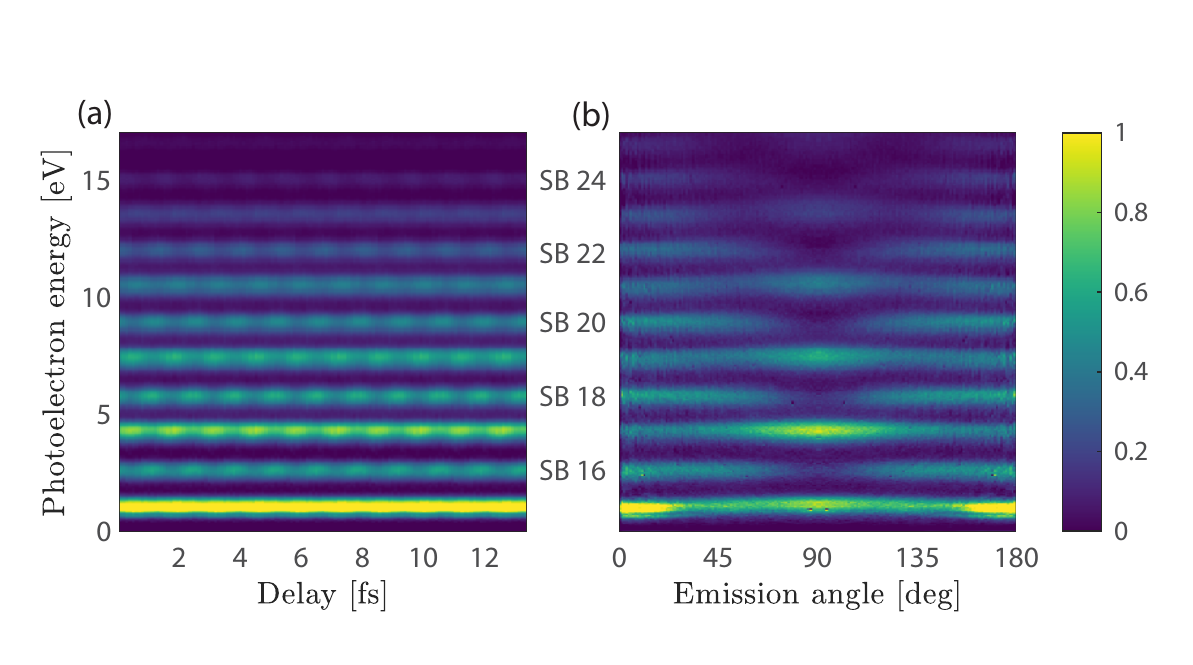}
     \includegraphics[width=0.9\textwidth]{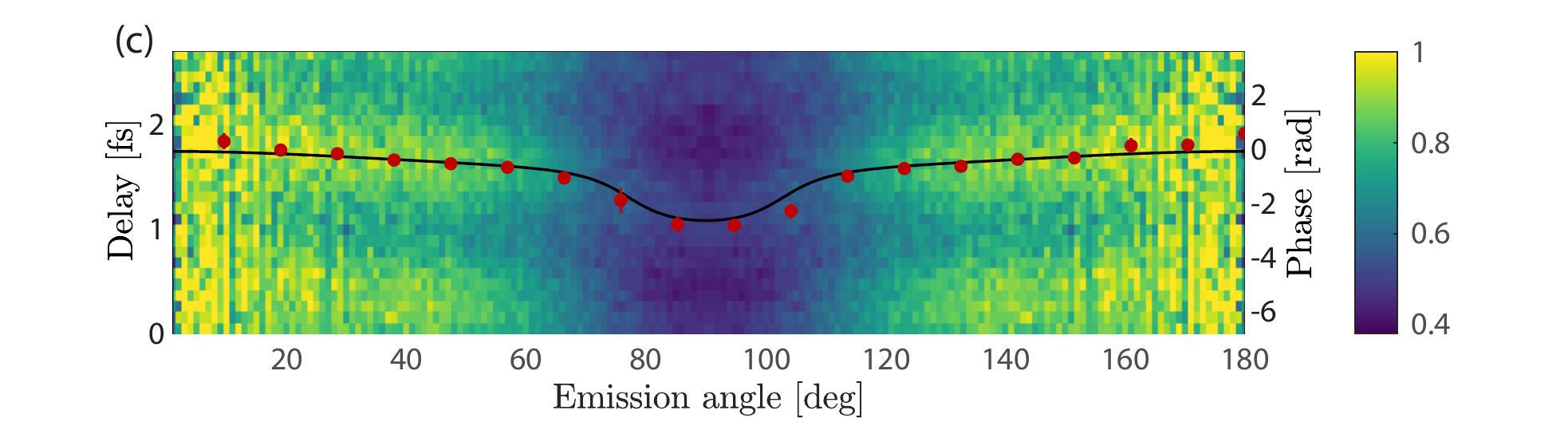}
  \caption{\label{fig2}\textbf{Angle and delay-dependence of photoelectron spectra:}  (a) Angle-integrated spectrum as a function of the delay; (b) Angle-resolved two-photon photoelectron spectrum integrated over all delays; (c) Energy-integrated sideband 16 as a function of angle and delay. The red dots show the extracted phase, and the black line shows the result of a simulation (see Methods section).}
\end{figure}

Figure 1 describes the principle of our experiment. Photoionization from the $2p^6$-ground state of neon leads to photoelectrons of different angular momenta $s$ or $d$. Additional absorption or emission of IR photons, through transitions between continuum states, called ``continuum-continuum'' (cc) transitions, leads to sidebands in the photoelectron spectrum. These sidebands can be reached by two interfering quantum paths (solid and dashed lines in Figure 1) \cite{PaulScience2001,VeniardPRA1996}. Our experiment consists in recording the photoelectron spectra as a function of delay between the XUV and IR pulses and emission angle (see experimental details in the Methods section).  

Figure 2 presents angular-integrated delay-dependent (a) as well as angular-resolved delay-integrated photoelectron spectra (b). It shows contributions from absorption of harmonics 15 to 25 as well as sidebands 16 to 24 (the number indicates the total energy absorbed in units of the IR photon energy), as a result of the interaction between released electrons and the infrared field. The sidebands in Figure 2(a) oscillate at frequency $2\omega$, where $\omega$ is the angular frequency of the driving laser. As shown in \cite{KlunderPRL2011}, the phase of the oscillation depends on the intrinsic group delay of the attosecond pulses as well as on a delay due to the photoionization process. The photoelectron peaks corresponding to absorption of harmonics oscillate with opposite phase, due to redistribution of the electrons from the main absorption peaks to the sidebands \cite{BertolinoPRR2021}. Figure 2(b) presents the angular dependence of the delay-integrated photoelectron signal. The sidebands are maximized at the angles of 0° and 180°, corresponding to emission along the common polarization axis of the XUV and IR fields. In contrast the main absorption peaks present a maximum at 90°. 

Finally, Figure 2(c) shows the delay and angle dependence of sideband 16. The oscillation phase is extracted by fitting a cosine to the temporal evolution for each angle. The red dots in Figure 2(c) show the measured phase as a function of the emission angle. 
The phase variation around 80° is close to 2 rad. The experimental data is compared to simulations (black line) based on calculations presented in the Methods section. 

The observed angular structure of the sidebands depends not only on the interference between the different partial waves of the angular momentum channels reached via single photon ionization, but is also strongly influenced by the cc-transitions~\cite{Heuser2016,Cirelli2018}. The additional interaction with the IR field leads to an increase in the number of angular channels (see Figure 1, where only the $m=0$ angular path is indicated), and modifies the radial amplitude and phase of the outgoing photoionization wavepacket \cite{DahlstromCP2013,Fuchs:20,Joseph_2020}.
In the following, we present a general method to retrieve the amplitude and phase of each photoionization angular channel from experimental data.

\begin{figure}[t!]
  \centering
     \includegraphics[width=1\textwidth]{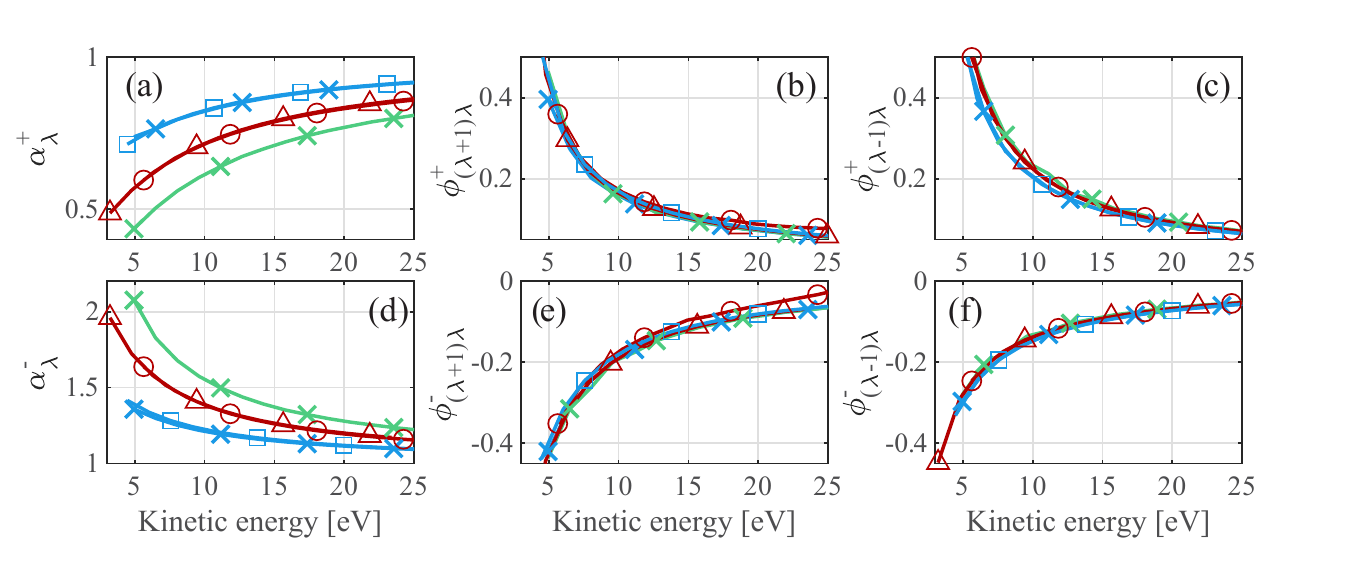}
  \caption{\label{fig3}\textbf{Continuum-continuum transitions:} Calculated amplitude ratios $\alpha_{\lambda \ell}^\pm$ (a,d) and continuum-continuum phases $\phi_{L \lambda}^{\pm}$ (b-c and e-f) for the absorption (a-c) and emission (d-f) processes. For the amplitude ratios, the curves correspond to different intermediate states ($\lambda$=1,2,3 in blue, red, green, respectively) and different atoms/initial states (square, He, 1s initial state); (cross, Kr, 3d); (triangle, Ne, 2p); (circle, Ar, 3p). (b,e) refer to transitions with increasing angular momentum, $L=\lambda+1$, while in (c,f), $L=\lambda-1$ .} 
\end{figure}

The angle- and delay-dependent sideband intensity can be written as

\begin{align}\label{Isb}
I_{SB}(\theta,\tau) \!\! \propto \!\! \int  \!\! d\phi \!\! \sum_{m=0,\pm 1} \!\! \Big|\sum_{L=1,3\atop{\lambda=0,2}} \!\! \mathcal{M}_{L\lambda \ell m}^{+}Y_{Lm}(\theta,\phi)\text{e}^{\text{i}\omega\tau} \!\! + \!\! \mathcal{M}_{L\lambda \ell m}^{-}Y_{Lm}(\theta,\phi)\text{e}^{-\text{i}\omega\tau}\Big|^2 \!\!,
\end{align}
where ($\pm$) refers to the pathways with IR absorption ($+$) or emission ($-$), 
$\mathcal{M}_{L\lambda \ell m}^{\pm}$ is the two-photon transition matrix element with final state angular momentum $L$, intermediate state orbital angular momentum $\lambda$, initial state orbital angular momentum $\ell$ ($\ell=1$ in the case studied here) and initial magnetic quantum number $m$, which is kept constant in the two-photon transition, since the XUV and IR fields are linearly polarized in the same direction. $Y_{Lm}(\theta,\phi)$ are spherical harmonics and $\text{e}^{\pm\text{i}\omega\tau}$ is the phase term introduced by absorption and emission of the IR photon with angular frequency $\omega$, which depends on the delay $\tau$ between the XUV and IR fields. The transition matrix element can be decomposed into amplitude and phase terms as
 \begin{equation}\label{ME}
     \mathcal{M}_{L\lambda \ell m}^{\pm}= C_{L\lambda}^mC_{\lambda \ell}^m~\sigma_{L\lambda \ell}^{\pm}~\text{e}^{\text{i}(\phi_{L\lambda}^{\pm}+\varphi_{\lambda \ell}^{\pm}+\Phi_{2q\mp1})}.
 \end{equation}
Here, $C_{L\lambda}^m, C_{\lambda \ell}^m$ are known angular coefficients (See Eq.~(S1) of the Supplementary Material, SM)
and $\sigma_{L\lambda \ell}^{\pm}$ is the radial amplitude. The phase term includes three different contributions: $\varphi_{\lambda \ell}^{\pm}$ is the phase associated to the one-photon ionization channel $\ell \rightarrow \lambda$, $\phi_{L\lambda}^{\pm}$ is the cc-phase, and $\Phi_{2q\mp 1}$ the phase of the $(2q\mp 1)^\mathrm{th}$ harmonic field. The characterization of multi-channel one-photon ionization using two-photon interferometry requires the determination of all these quantities, from either theoretical arguments or experimental measurements.

Applying and extending the results of previous work \cite{DahlstromCP2013,Fuchs:20,Busto2019}, we use the universal behavior of the cc-transitions to determine some of the terms in Eq.~(\ref{ME}), thus reducing the number of unknown quantities.
 Figure \ref{fig3} (a,d) presents the ratios between the two-photon transition radial amplitudes, calculated as described in the methods section and defined as
\begin{equation}
 \alpha_{\lambda \ell}^{\pm} = \frac{\sigma_{(\lambda-1)\lambda \ell}^{\pm}}{\sigma_{(\lambda+1)\lambda \ell}^{\pm}},
\end{equation}
for increasing or decreasing angular momentum from the same intermediate state as the function of the kinetic energy of the electron. The different curves correspond to different intermediate states ($\lambda$=1,2,3 in blue, red and green) and different atoms and initial states. The ratio shows an universal behavior~\cite{Busto2019}, independent of the atom. Only the orbital angular momentum of the intermediate state, $\lambda$, is of importance.

We present $\phi_{L\lambda}^{\pm}$ in Figure \ref{fig3} for increasing (b,e) and decreasing (c,f) angular momenta, in the absorption (b,c) and emission cases (e,f). The colors and symbols, corresponding to different intermediate angular momenta and atoms/initial states are indicated in the figure caption. These results show that the variation of the continuum-continuum phase is universal, depending mainly on whether the IR photon is absorbed or emitted. For the absorption process, the cc-phase decreases as a function of kinetic energy and is positive, while for the emission, it increases and is negative. Note that the phases are not mirror image of each other, i.e. $\phi_{L\lambda}^{+} \neq -\phi_{L\lambda}^{-}$.  The cc-phases depend on whether the angular momentum increases or decreases, especially at low kinetic energy, as observed by comparing  (b) and (c), or (e) and (f). Finally, the cc-phases depend only slightly on the intermediate angular momentum (compare blue and red curves) and not at all on the atomic system (e.g. compare circle and triangle) in the range of energies studied here. 

Our channel-resolved amplitude and phase retrieval is based on the knowledge of the cc-transitions. The unknown and known quantities in Eq.~(1), after having expressed the transition matrix elements as in Eq.~(2), and used the available information in Fig.~\ref{fig3}, are indicated in Table 1 for an initial $\ell=1$ state. We note that only the phase difference $\Delta\Phi_{2q}=\Phi_{2q+1}-\Phi_{2q-1}$, and not the individual high-order harmonic phases, plays a role in Eq.~(1). The number of unknown quantities is therefore nine. While here the case of $\ell=1$ is shown, this is also true for higher initial angular momenta.

\begin{table}[h!]
\begin{center}
 \caption{\label{tab1} Unknown and known quantities involved in Eq.~(1)}
\begin{tabular}{ m{1.8cm} | x{3.3cm} | x{4.5cm} }
 & Unknown  & Known  \\ 
 &  quantities &  quantities  \\ 
 \hline\hline
 \vspace{0.2cm}
 Atomic phases & $\varphi_{01}^\pm$, $\varphi_{21}^\pm$ & $\phi_{10}^\pm$, $\phi_{12}^\pm$, $\phi_{32}^\pm$, {\scriptsize from Fig. \ref{fig3}(b,c,e,f)}\\
 \vspace{0.2cm}
 2-photon amplitudes & $\sigma_{101}^\pm$, $\sigma_{321}^\pm$ & $\alpha_{21}^\pm$, {\scriptsize from Fig. \ref{fig3}(a,d)}\\
 & & $\rightarrow~\sigma_{121}^\pm=\alpha_{21}^\pm\sigma_{321}^\pm$\\
 \vspace{0.2cm}
 Harmonic phase & $\Delta\Phi_{2q}=\Phi_{2q-1}-\Phi_{2q+1}$ & \\
\end{tabular}
\end{center}
\end{table}

\begin{figure}[t!]
  \centering
  \includegraphics[width=1\textwidth]{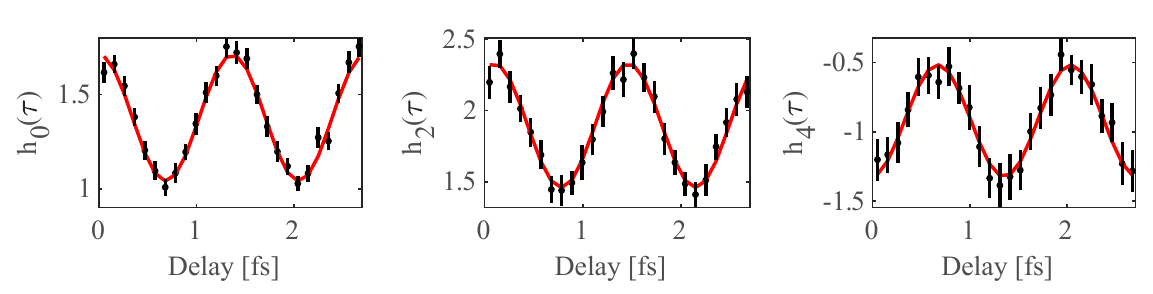}
  \caption{\label{h_func_v1}\textbf{Global fit to experimental data: } Delay dependence of the coefficients $h_i(\tau)$, $i=0,2,4$ for sideband 18. The black dots are obtained from the angular distributions (Eq.~\ref{Isb_le}) for each delay and the error bars correspond to one standard deviation. The red curves are the result of a simultaneous fit of the delay dependent $h_i$ functions using Eqs.~(S2-S4) in the SM. }
\end{figure}

We determine these nine unknown quantities using a global fit to our experimental measurements. In general, multiphoton electron angular distributions can be written as an expansion in Legendre polynomials~ \cite{Fuchs:20,Joseph_2020,ReidARPC2003,ArnousPRA1972}. For a two-photon transition, without parity mixing, the expansion needs only three polynomials, $P_0(x)=1$, $P_2(x)=(3x^2-1)/2$, and $P_4(x)=(35x^4-30x^2+3)/8$, reading as 
\begin{equation}\label{Isb_le}
    I_{\text{SB}}(\theta,\tau)=h_0(\tau)+h_2(\tau)P_2(\text{cos}~\theta)+h_4(\tau)P_4(\text{cos}~\theta). 
\end{equation}
The theoretical expressions for the coefficients $h_i(\tau)$, $i=0,2,4$, can be obtained by expanding Eq.~(\ref{Isb}) and replacing the products of spherical harmonics by Legendre polynomials, leading to Eqs.~(S2-S4) of the SM. Figure \ref{h_func_v1} shows the variation of $h_i(\tau)$ extracted from the experimental data for each delay (black points).
As is clear from Eqs.~(S2-S4) in the SM and from Figure \ref{h_func_v1}, each $h_i(\tau)$ oscillates with the delay $\tau$ at the frequency $2\omega$ and is therefore fully determined by three quantities: mean value, amplitude and phase. 
Thus, a total of nine parameters describe the angle and delay dependence of the sideband signal $I_{\text{SB}}(\theta,\tau)$. This implies that the nine unknown quantities in table~\ref{tab1} can be determined through a global fit of the three analytical expressions of $h_i(\tau)$ given in the SM, to the experimentally measured coefficients in Figure~\ref{h_func_v1}. The result of such a global fit is shown by the red lines in Figure \ref{h_func_v1}. 

\begin{figure}[t!]
  \centering
  \includegraphics[width=0.9\textwidth]{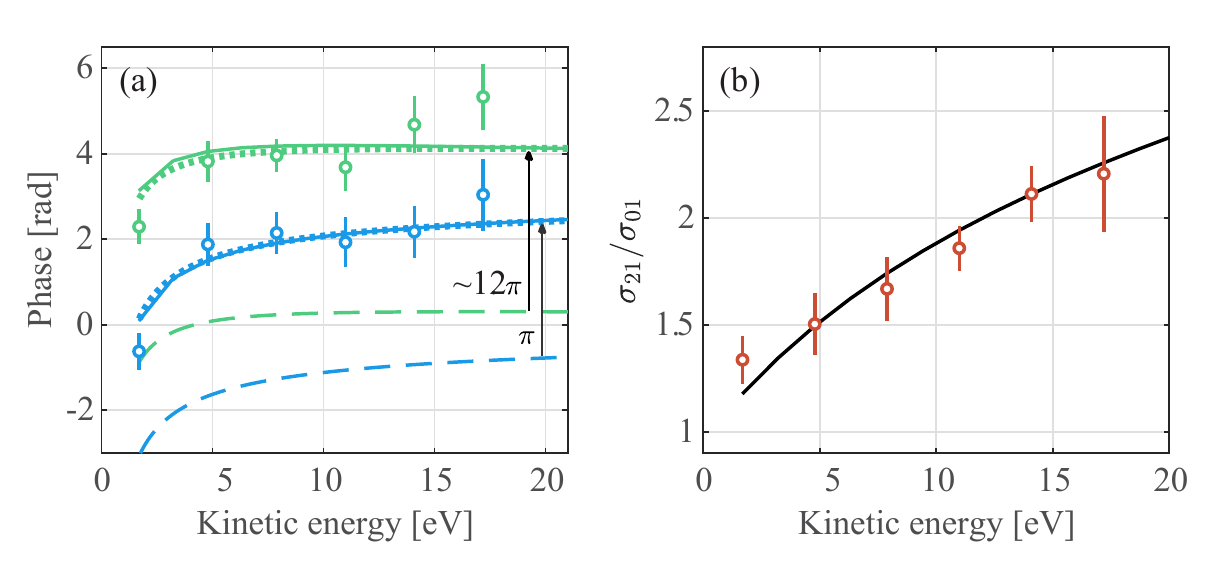}
  \caption{\label{fig5}\textbf{Extracted one-photon amplitudes and phases as a function of the kinetic energy.} (a) Scattering phases $\varphi_{01}$ (green) and $\varphi_{21}$ (blue) for experimental (dots) and simulated data (solid lines). The dashed lines correspond to the contributions from the Coulomb phase $\varsigma_{0}$ (green) and  $\varsigma_{2}$ (blue), where for the dotted lines the effect of the short-range potential and the centrifugal effect is added: $\varsigma_\lambda+\delta_\lambda-\pi\lambda/2$. For $\lambda=0$ this shift is dominated by the short range potential ($\delta_0\approx 1.2\pi$) and for $\lambda=2$ by the centrifugal effect ($\pi$), as indicated by the arrows. (b) Ratio between amplitudes $\sigma_{21}$ and $\sigma_{01}$ extracted from XUV only data (dots). The black line shows the result of angular-channel-resolved many-body perturbation simulation.} 
\end{figure}

The nine unknown quantities in table~\ref{tab1} are not completely independent, since the one-photon ionization phases only appear as differences in Eqs.~(S2-S4) (SM). We therefore lock $\varphi_{01}^-$ for the first sideband and determine all of the others ($\varphi_{01}^+$,$\varphi_{21}^{\pm}$). 
To map out the energy dependence of the one-photon phases, the global fit is repeated for each sideband. Using the fact that  $\varphi_{\lambda 1}^-(SB_n)=\varphi_{\lambda 1}^+(SB_{n+2})$, we iteratively retrieve the one-photon phases, as a function of energy, for the two possible angular momenta. From now on, we drop the $\pm$ superscript to describe the one-photon phase, as it only refers to the path used for the determination. The experimental results for $\varphi_{01}$ and $\varphi_{21}$ are shown in Fig.~\ref{fig5}(a) in green and blue symbols respectively, together with the phases calculated using many-body perturbation theory, as described in the methods section (solid lines).

The phase $\varphi_{\lambda \ell}$ can be written  as the sum of the scattering phase $\eta_\lambda$ and a contribution from the centrifugal barrier $-\pi \lambda/2$~\cite{DahlstromCP2013}. The scattering phase is itself the sum of the Coulomb phase $\varsigma_{\lambda}=\arg \Gamma(\lambda+1-iZ/k)$ and a contribution $\delta_\lambda$ from the short range potential. (Here $Z$ is the atomic number, $k=\sqrt{2mE}/\hbar$ the wavenumber and $\Gamma$, the gamma function). In order to emphasize the influence of the short range potential, we also show in Figure~\ref{fig5}(a) $\varsigma_0$ and $\varsigma_2$ (green and blue dashed lines), as well as $\varsigma_0+\delta_0$ and $\varsigma_2+\delta_2-\pi$ (green and blue dot-dashed lines), where $\delta_\lambda$ are taken from Kennedy and Manson~\cite{KennedyPRA1972}. The calculated phases are very close to $\varsigma_\lambda+\delta_\lambda-\pi\lambda/2$ (compare solid and dot-dashed lines). 
The difference between $\varphi_{01}$ and $\varsigma_0$ is due to the short range potential 
contributing by  $\sim 1.2\pi$, while that between $\varphi_{02}$ and $\varsigma_2$ is mainly due to the effect of the centrifugal barrier, leading to a $\pi$ phase shift. For $\lambda=2$, short range effects are small due to the centrifugal barrier that prevents the electron to come close to the core. The measured phases are also in excellent agreement with the predictions from quantum defect theory at threshold, leading to $\delta_\lambda=\pi \mu_\lambda \simeq 1.3\pi$ for $\lambda=2$ and $\delta_\lambda \simeq 0$ for $\lambda=0$, ($\mu_\lambda$ is the quantum defect extracted from experimental energy values) \cite{FriedrichBOOK}. 
Since the short range effects for the $s$-electron are comparable to the effect of the centrifugal barrier for the $d$-electron, the difference between $\varphi_{01}$ and $\varphi_{21}$ therefore mostly reflects the difference in Coulomb phases. The increase of the phases with energy, which is similar for the two angular momenta also follows the behavior of the Coulomb phases.

As explained above, our procedure allows us to determine the phases as function of energy and angular momentum up to a global phase offset.  After adjustment of this phase offset, the theoretical and experimental results are found to be in excellent agreement. Both the energy dependence and the difference between the angular channels is well reproduced by the experimental results. The observed deviation for the highest energy might be due to low statistics originating from the low intensity of the 25$^\mathrm{th}$ harmonic.

Finally, in order to fully characterize single photon ionization, it is necessary to also determine the one-photon amplitudes $\sigma_{\lambda\ell}$. However, it is not possible to extract $\sigma_{\lambda\ell}$ from the two-photon ionization data. Instead, we perform additional measurements in the absence of the IR field, in order to access directly the one-photon transition amplitudes. The one-photon angle-dependent photoelectron signal is written as:
\begin{equation}\label{OP_signal}
I_{\text{H}}(\theta)\propto \int d\phi~\sum_{m=0,\pm 1}\Big|\sum_{\lambda=0,2}M_{\lambda\ell m}Y_{\lambda m}(\theta,\phi)\Big|^2.
\end{equation}
Here, $M_{\lambda\ell m}$ is the one-photon matrix element from the ground state with angular momentum $\ell$ and magnetic quantum number $m$ to the continuum state with angular momentum $\lambda$, which can be written as
\begin{equation}
\label{OP_ME}
M_{\lambda\ell m}\approx C_{\lambda\ell}^m~\sigma_{\lambda\ell}~\text{e}^{\text{i}(\varphi_{\lambda\ell}+\Phi_{2q+1})},
\end{equation} 
where $2q+1$ is the order of the harmonic used for the photoionization. $I_{\text{H}}(\theta)$ can be written as an expansion of Legendre polynomials $P_0$ and $P_2$ [see Eqs. (S5) and (S6) in the SM]. The coefficients of the expansion $h_0$ and $h_2$ are extracted from the experimental data. Using the one-photon phases obtained previously, we determine the relative radial amplitudes of the $\lambda=0$ and $\lambda=2$ channels. The ratio between these amplitudes is shown in Figure \ref{fig5}(b), together with the calculated one. 
The ratio is above one for all energies, thus in agreement with Fano's propensity rule for one-photon absorption.


The retrieval of the amplitudes and phases of angular momentum channels in photoionization has been a major challenge during the last decades, and has only been realized for specially prepared atoms or for single angular channels. Here, we present a method using laser-assisted photoionization, based on the universality of the continuum-continuum transitions to retrieve the one-photon photoionization amplitudes and phases for the different angular channels. We apply the method both experimentally and numerically to characterize photoionization from the 2p$^6$ shell of neon. Our method is general and can be applied to other shells (e.g. $d$) and thus to more complex atomic systems. In particular, the study of the phases of the different angular momentum channels, unravels the interplay between short-range, correlation and/or centrifugal effects.

\section{Acknowledgments}

The authors acknowledge support from the Swedish Research Council, the European Research Council (advanced grant QPAP), the Knut and Alice Wallenberg Foundation, and the Crafoord Foundation. 

\clearpage
\section*{Methods}

\begin{figure}[t!]
  \centering
     \includegraphics[width=0.7\textwidth]{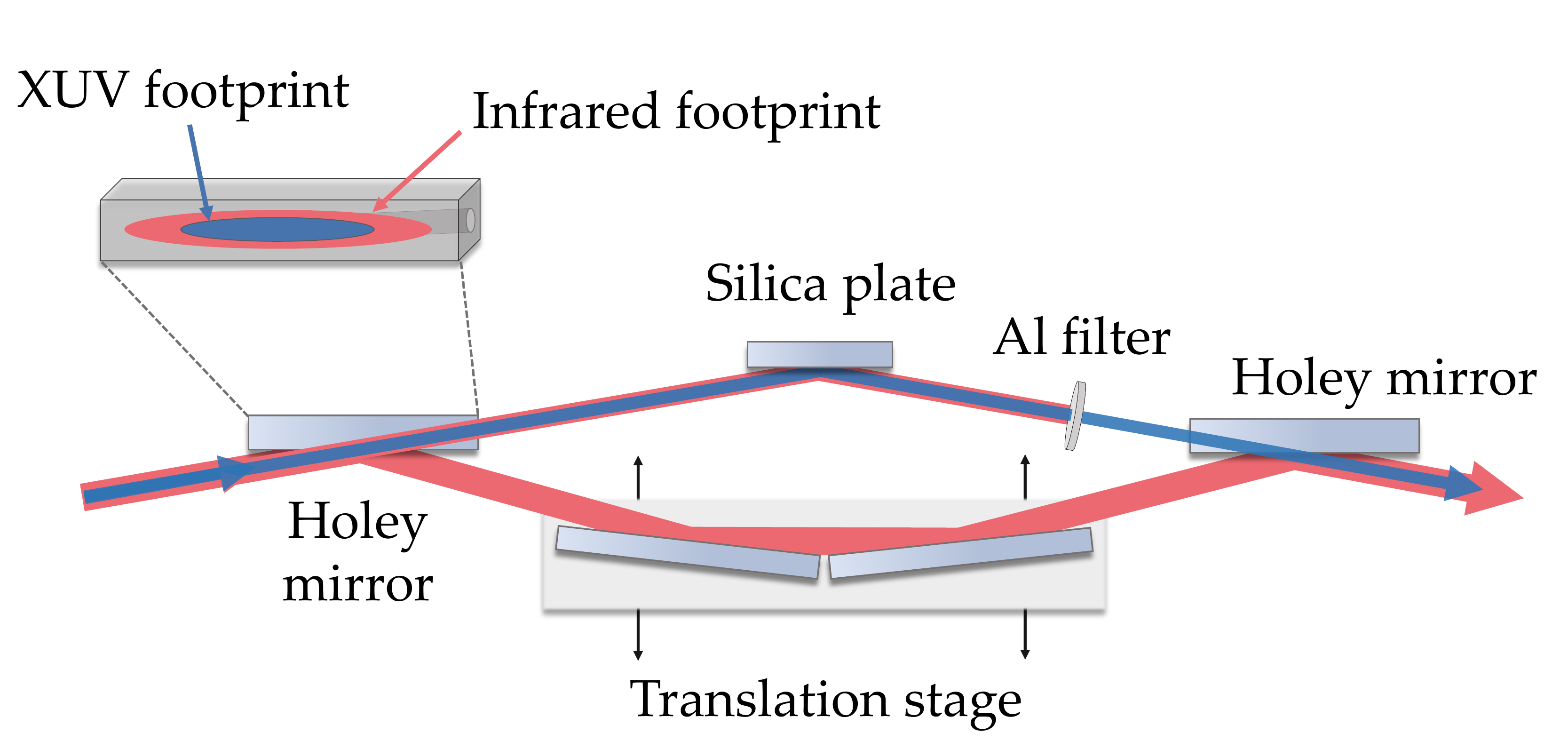}
  \caption{\label{interferometer}\textbf{Schematic drawing of the XUV-infrared interferometer.} Two holey mirrors split and recombine the XUV and infrared components of the beam and a linear translation stage introduces a time delay between them.}
\end{figure}

The XUV attosecond pulse trains (ATPs) used in the experiments are synthesized using high-order harmonic generation (HHG) by focusing 40 fs, 806 nm, $\sim$45 mJ pulses from a 10 Hz Ti:sapphire laser into a pulsed gas jet of Argon in a loose focusing geometry ($\sim$8.7 m focal length) \cite{ManschwetusPRA2016}. The obtained XUV spectrum spans from $\sim$20 to 45 eV. The beamline is designed to generate high-flux high-order harmonics, hence the loose focusing geometry \cite{RudawskiRSI2013}.

The APTs are separated from the fundamental infrared field by taking advantage of the lower divergence of the XUV beam.  Figure \ref{interferometer} shows a scheme of the newly developed interferometer, which consists of a set of holey mirrors creating two interferometric arms. The hole diameter is chosen in a way such that the entire XUV beam propagates through the hole, whereas the reflected part only consists of fundamental infrared radiation. A translation stage in the infrared arm varies the path difference between both arms and hence introduces a time delay. A 200 nm thick aluminum filter in the XUV arm blocks the remaining infrared and the intense low-order harmonics. Due to the long beamline design, traditional interferometers add the risk of pointing instabilities and temporal jitters, while propagating over such long distances. Hence, we designed this compact in-line interferometer, such that the path difference is kept as short as possible, which results in excellent temporal and spatial stability. 

Both beams, after being collinearly overlapped, are focused tightly using two toroidal mirrors in a Wolter configuration \cite{Coudert:17}. In focus they interact with the neon target gas introduced using a pulsed Even-Lavie valve \cite{Even2014} and a set of two skimmers. The resulting photoelectrons are detected by a velocity map imaging spectrometer (VMIS) with the ability to record angle-resolved momentum distribution \cite{EppinkRSI1997,Rading2018}. The 2D projections of the momentum distribution of the photoelectrons is recorded by a CCD camera imaging a phosphor screen coupled to a set of multi-channel plates. Since the VMIS is recording the 2D projection of the 3D momentum distribution, an inverse Abel transform has to be applied, which is done using an iterative method \cite{VrakkingRSI2001}.

Our calculations are based on a one-electron Hamiltonian, with a Dirac-Fock potential plus a correction that ensures the correct long-range potential for ionized photoelectrons \cite{DahlstromPRA2012}. The absorption of one ionizing photon is treated within the Relativistic
Random Phase Approximation with Exchange (RPAE) resulting in a so-called perturbed wave function
describing the ionized electron. The method accounts for  important many-body effects such as
inter-channel coupling and  ground-state correlation. Exterior complex scaling is used in order
to be able to use a finite numerical box.

The complex-valued two-photon matrix elements, expressed in Eq. (\ref{ME}),  are then calculated as the transition from the perturbed wave function to the final continuum state in each angular momentum channel, following the procedure described in \cite{DahlstromPRA2012,DahlstromJPB2014,VinbladhPRA2019} for the non-relativistic case. The integration is performed numerically out to a distance far outside the atomic core, but within the unscaled region, while the last part of the integral is carried out using analytical Coulomb waves along the imaginary radial axis. The amplitude and phase shift of these Coulomb waves are determined from the numerical solutions for the perturbed wave function and for the final state describing a free electron within the potential of the remaining ion. The numerical stability is monitored by comparison of different “break points” between the numerical and analytical descriptions.

\bibliographystyle{unsrt}
\bibliography{Ref_lib}

\vfill

\end{document}